\documentclass[10pt,letterpaper]{article}

\usepackage[final=true]{cogsci}

\usepackage[
  style=apa,
  natbib=true,
  annotation=false,
]{biblatex}
\usepackage{float} 
\usepackage{placeins}
\usepackage{stfloats}

\usepackage{tikz}
\usetikzlibrary{positioning}

\usepackage{graphicx}
\usepackage{amsmath}
\usepackage{bm}
\usepackage{booktabs}
\usepackage{xcolor}
\usepackage{colortbl}

\title{Popularity Feedback Constrains Innovation in Cultural Markets}

\author[1]{\mbox{Lucas Gautheron}}
\author[2]{\mbox{Raja Marjieh}}
\author[3]{\mbox{Dalton C. Conley}}
\author[4]{\mbox{Seth Frey}}
\author[5]{\mbox{Hannah Rubin}}
\author[5]{\mbox{Mike D. Schneider}}
\author[6]{\mbox{Ofer Tchernichovski}}
\author[7,8]{\mbox{Nori Jacoby}}
\affil[1]{Evolution, Science and Society, University of Missouri, Columbia, Missouri, United States
\texttt{(lucas.gautheron@gmail.com)}}
\affil[2]{Department of Psychology, Princeton University, Princeton, New Jersey, United States}
\affil[3]{Department of Sociology, Princeton University, Princeton, New Jersey, United States}
\affil[4]{Department of Communication, UC Davis, Davis, California, United States}
\affil[5]{Department of Philosophy, University of Missouri, Columbia, Missouri, United States}
\affil[6]{Department of Psychology, Hunter College, New York, New York, United States}
\affil[7]{Cornell Computational Cognition Lab, Cornell University, Ithaca, New York, United States}
\affil[8]{Max Planck Institute for Empirical Aesthetics, Frankfurt, Germany}

\begin{document}

\maketitle

\begin{abstract}

Real-world creative processes ranging from art to science rely on social feedback-loops between selection and creation. Yet, the effects of popularity feedback on collective creativity remain poorly understood. We investigate how popularity ratings influence cultural dynamics in a large-scale online experiment where participants ($N = 1\,008$) iteratively \textit{select} images from evolving markets and \textit{produce} their own modifications. Results show that exposing the popularity of images reduces cultural diversity and slows innovation, delaying aesthetic improvements. Popularity feedback is associated with changes to both selection and creative stages. During selection, popularity information triggers cumulative advantage, with participants preferentially building upon popular images, reducing diversity. During creation, participants make less disruptive changes, and are more likely to expand existing visual patterns. Feedback loops in cultural markets thus not only shape selection, but also, directly or indirectly, the form and direction of cultural innovation.

\end{abstract}

\section{Introduction}

In many domains of human culture such as science, arts, or social media, producers of cultural goods can observe and compare the popularity of existing cultural variants. These popularity signals can trigger a form of \textit{popularity feedback} by influencing later cultural innovations. In science, for instance, popularity information is widely available in the form of citation counts, which are known to exhibit cumulative advantage as scientists preferably cite highly popular papers \citep{barabasi1999emergence,Wang2013}. On the producer side, disproportionate attraction to popular ideas may diminish intellectual diversity, distort assessments of the long-term potential of certain research directions, and alter the space of possibilities that scientists are willing to explore \citep{Kummerfeld2016}. Alternatively, popularity feedback may enhance social learning by providing reliable fitness signals \citep{Miu2018}. Additionally, popularity feedback may alter innovation strategies in ways unappreciated by blind variation models \citep{Campbell1960}, altering the shape of cultural products. For instance, in artistic domains, we can hypothesize that access to popularity feedback  encourages producers to amplify or exaggerate arbitrary features present in successful cultural goods. This could create a ``cultural runaway'' \citep{boyd1988culture} analogue to biological runaway selection, an evolutionary process responsible for exaggerated ornaments across several species and driven by a positive feedback between a trait and preference for this trait \citep{Lande1981,Lehtonen2012,prum2012aesthetic}. Similarly, digital platforms such as social media may be influenced by popularity feedback via shares or likes. While no evidence of such popularity biases was found in Twitter \citep{Carrignon2019,Youngblood2023}, these could arise in platforms with higher stakes for producers.

Despite its potential significance, few works have addressed the evolutionary consequences of popularity feedback. 
Prior research has often neglected the incidence of feedback loops on the form and direction of cultural innovations, owing to a stronger focus on selection in cultural evolutionary literature \citep{Andr2023}. Better studied feedback mechanisms include frequency-dependent selection, when the adoption rate of cultural variants depends on their incidence among a population, due to e.g. conformity biases \citep{Acerbi2014,Newberry2022}, or varying utility, as in the case of ``network goods'' \citep{Bjrkegren2018}. A related phenomenon is cumulative advantage, when popular cultural goods that are already popular attract more attention, independently from their intrinsic worth. For instance, \citet{Salganik2006} have shown that access to popularity information in artificial markets of songs ultimately led to more unequal and unpredictable market shares. By contrast, the effects of such feedback loops in markets involving both consumers \textit{and} producers of cultural goods has been far less studied experimentally, with the exception of \citet{Balietti2016}, who demonstrated that competitive incentives influence peer-review feedback processes involving both creators and reviewers. Yet, this study did not consider social feedback via popularity information, and it did not directly examine how social information affects the shape of cultural outputs. 

We address these gaps in a large-scale online experiment in which participants ($N=1\,008$) iteratively observe a cultural market of images and populate it with their own drawing after having selected an image to improve. Our experiment thus implements both selection and creation steps, therefore capturing the two major forces of cultural evolution \citep{Mesoudi2021}. We find that access to popularity information (about how many times items were copied) leads to less diverse cultural markets; slower innovation; and initially less aesthetic and creative images, although this trend may be reversed in the long-run, possibly due to social feedback protecting high-quality innovations. These effects are associated with  changes to both selection and creation steps. During selection, popularity information triggers a process of cumulative advantage, whereby participants preferentially choose to build upon popular ideas, which undermines diversity and amplifies popularity inequalities. Finally, individuals make less disruptive changes under popularity feedback. We discuss what can possibly lead to these changes in the emergent cultural dynamics.

\section{Methods}

\begin{figure}[!h]
  \centering
  \begin{tikzpicture}
      \node[draw] (selection) {\includegraphics[width=4.8cm]{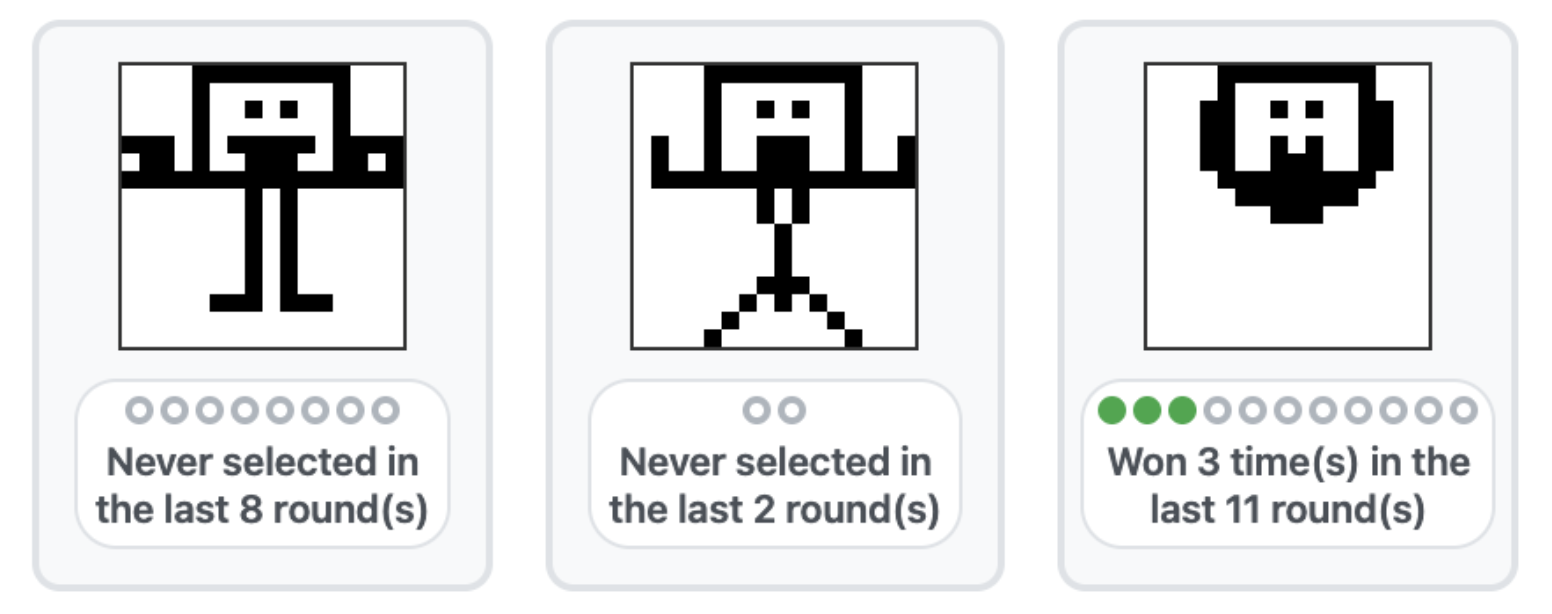}};

      \node[draw,right=0.7cm of selection] (creation) {\includegraphics[width=2.7cm]{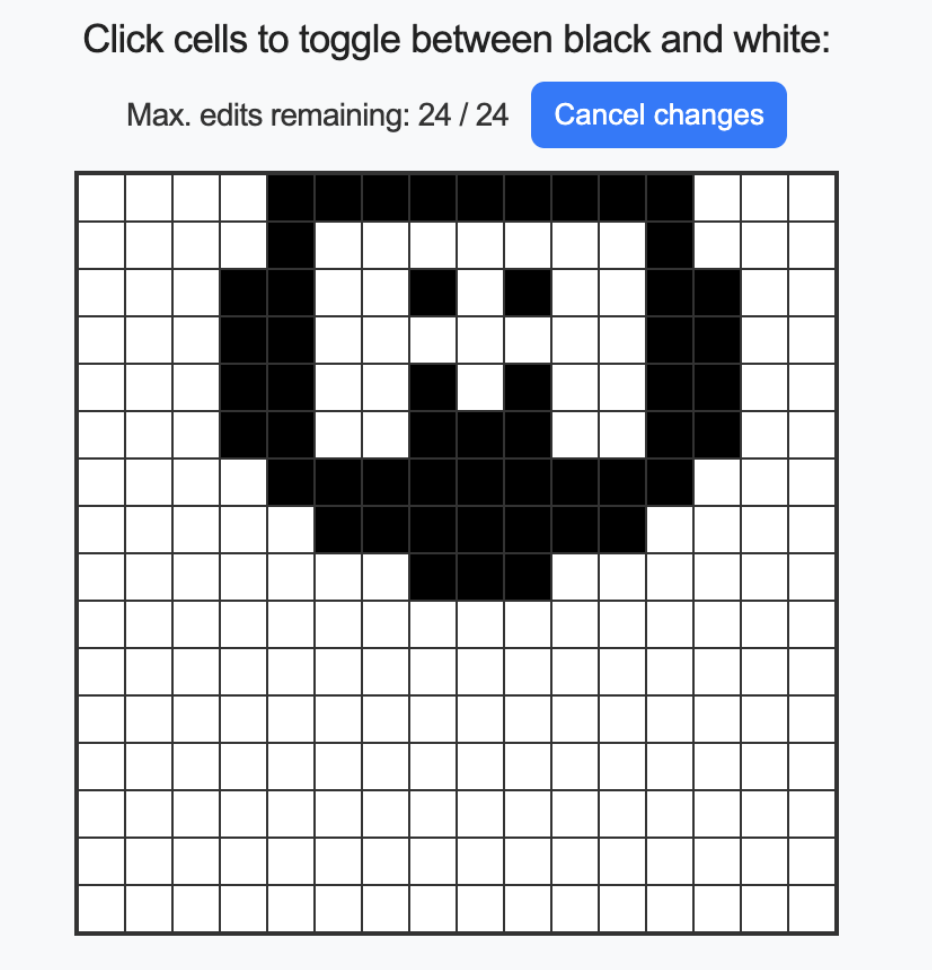}};
      
      \draw[->, thick, bend left=35] (selection.north) to node[above=0.2cm, midway, align=center] {Selection\\(Choose from the market)} (creation.north);
      
      \draw[->, thick, bend right=-35] (creation.south) to node[below=0.2cm, midway, align=center] {Creation\\(Add to the market)} (selection.south);
  \end{tikzpicture}
  \caption{Selection and creation steps. Participants observe a market of 12 images, choose one image, change $\geq 1$ and $\leq 24$ pixels, and add their own creation to the market.}
  \label{fig:process}
\end{figure}

We develop a large-scale collective experiment in which participants observe collections of images (16$\times$16 grids with black or white pixels), select one image to modify, and contribute their creation to the collection (Fig. \ref{fig:process}). The editing interface was similar to other pixel-based drawing experiments in the literature \citep{hart2017creative,kumar2022using, tchernichovski2025constraints}.
In the treatment group, during the selection step, participants observe the popularity of each item in the market (i.e. how many times the item had been previously selected to be edited). Participants are asked to behave so as to maximize the chance that their own work is selected by subsequent participants. This creates a strategic tension: while participants may choose to capitalize on ideas that have proven successful in the past, doing so implies competing with many similar images for attention, decreasing the chance of being selected. This dilemma thus invites either conformist strategies (building on popular trends, i.e. positive feedback) or anti-conformist approaches (pursuing novel ideas, i.e. negative feedback). The process mirrors dynamics in many creative cultural domains, including science, where researchers must balance engaging with trending topics against differentiating their contributions. After selecting an image, participants are allowed to change no less than 1 and up to 24 pixels. This constraint ensures that participants engage with the selected image and thus increases the strategic importance of the selection step. In addition, the editing restriction dilutes the relative effect of individual contributions. 

\begin{figure*}[p]
    \centering
    \includegraphics[width=1\linewidth,trim={2.5cm 12.1cm 1.8cm 1.75cm},clip]{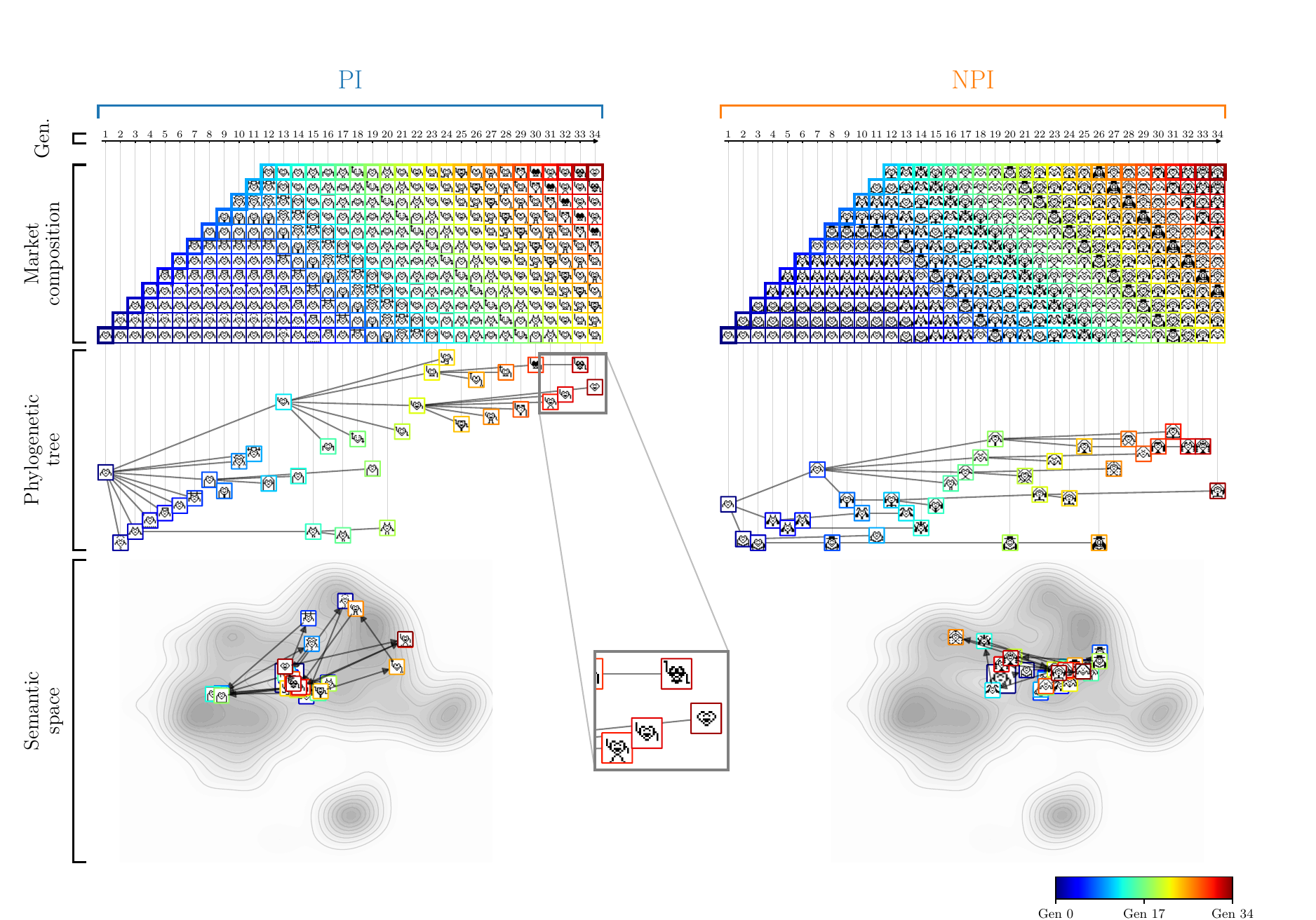}
    \caption{Evolution of two chains starting from the same image, over the first 35 generations, under conditions with popularity information (PI, to the left) and without (NPI, to the right). \textbf{Market composition}: at each generation (\textit{gen.}), participants observe a market containing the 12 latest creations in the chain (each column represents the contents of the market at a given generation). \textbf{Phylogenetic tree}: ancestry relationships between images can be represented in the form of a tree, from the first image (to the left, $t=1$) to the last (on the right, $t=34$).  \textbf{Semantic space}: images are assigned positions in a high-dimensional space using a machine learning vision model \citep{blip}, revealing the motion of each chain in their broader cultural landscape.}
    \label{fig:evolution}
\end{figure*}

We consider 128 isolated markets, initialized with a single image each, and evolving as independent chains over 60 generations ($7\,680$ images overall). For every chain, at any generation $g$, one participant observes a market of up to $T=12$ images produced between generations $g-1$ and $g-T$. Chains are divided into two conditions: the popularity information condition (PI), in which participants observe the popularity of each image in the market during the selection step (i.e., how many times the image was already selected and edited by previous participants); and the no-popularity information condition (NPI), in which this feedback is absent. Every chain from either condition is paired with a chain from the other condition, initialized with the same image (as in Fig. \ref{fig:evolution}).
Participants contributed only once for each chain and up to 8 chains overall, all from the same condition (fewer if they dropped out before the end of the experiment). 

The experiment was deployed online using PsyNet  \citep{harrison2020gibbs}. It includes $N=2\,015$ participants split between the cultural market experiment itself ($N=1\,008$) and additional image rating or labeling experiments ($N=1\,007$). Participants were recruited in the United States via Prolific, provided informed consent under a Cornell University-approved protocol (IRB0148995) or a Max Planck Society-approved protocol (2021-4/2024-36), and were compensated at a rate of approximately \$12 per hour.

\section{Results}

\subsection{Access to popularity information undermines cultural diversity and constrains innovation}

\begin{figure*}
    \centering
    \begin{minipage}{0.48\textwidth}
        \centering
        \includegraphics[width=\linewidth]{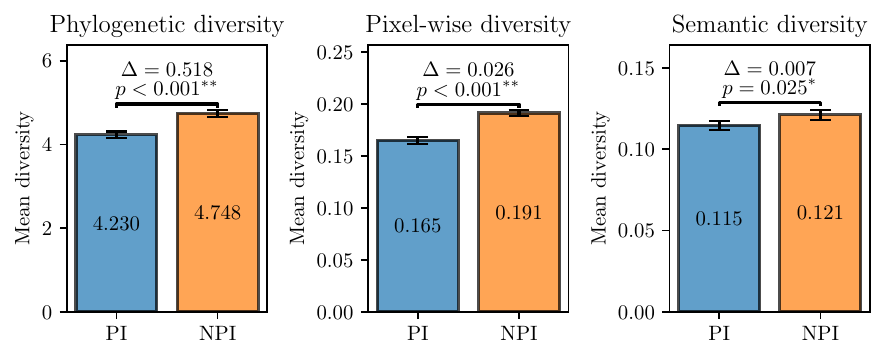}
        \caption{Average market diversity in the conditions with popularity information (PI) and without (NPI). Diversity is measured as the average distance between images in a market at any generation. Error bars indicate standard errors ($\pm 1$SE). Statistical significance is evaluated via a non-parametric permutation test pairing chains with identical initial images.}
        \label{fig:diversity}
    \end{minipage}
    \hfill
    \begin{minipage}{0.48\textwidth}
        \centering
        \includegraphics[width=\linewidth]{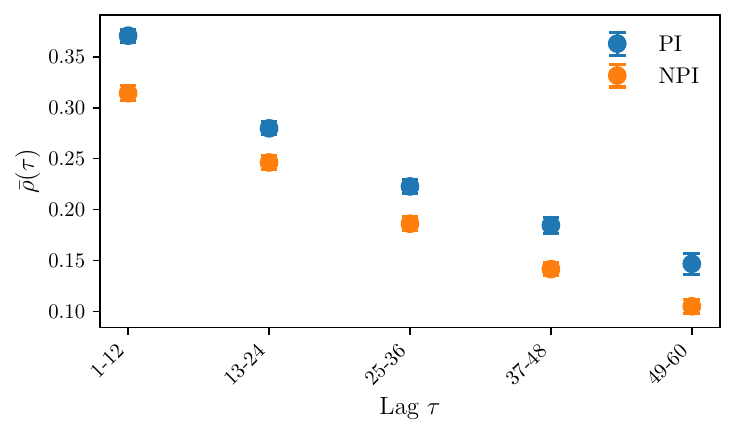}
        \caption{Average chain autocorrelation in semantic space $\bar{\rho}(\tau)$ under the conditions with popularity information (PI) and without (NPI) and for different time-lags $\tau$.}
        \label{fig:autocorrelation}
    \end{minipage}
\end{figure*}

We examine the consequences of popularity information for cultural diversity and innovation. First, at each generation of every chain, we measure cultural diversity as the average pairwise distance between images in the market, according to three distance measures: the phylogenetic distance (i.e., the degree of separation between images in their evolutionary tree); the pixel-wise (or hamming) distance, measuring the fraction of pixels that differ between two images; and the semantic distance between images (i.e., the cosine distance between high-dimensional image-embeddings computed by a vision-transformer model; \citealt{blip}). This reveals a statistically significant reduction in diversity across the three measures when social feedback is provided (Fig. \ref{fig:diversity}). 

\begin{figure*}[p]
    \centering\includegraphics[width=0.95\linewidth]{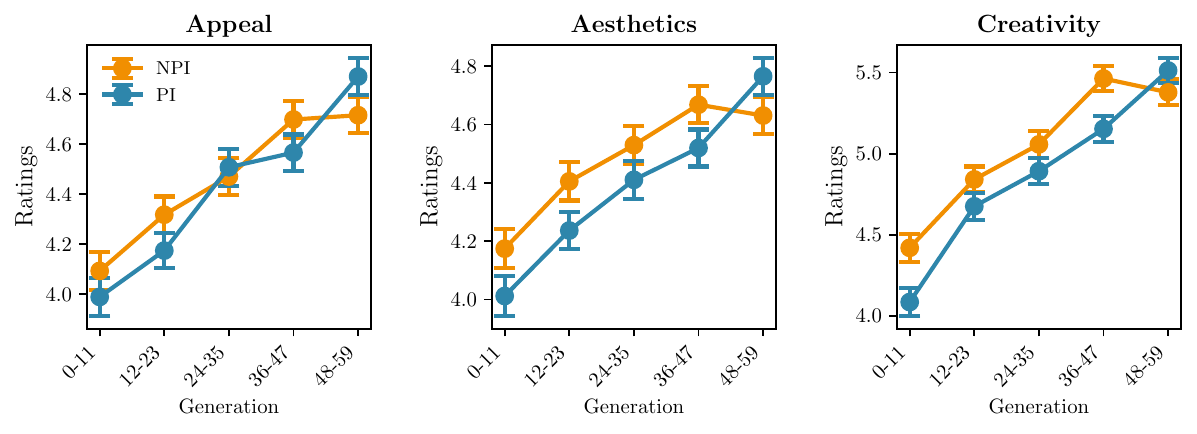}
    \caption{Quality of images throughout generations and across conditions. Error bars indicate standard errors  ($\pm 1$SE) on the chain averages (effective sample-size $n=64$).}
    \label{fig:ratings}
\end{figure*}

Popularity information may affect not only the state of cultural markets at any point in time, but also the rate of cultural innovation across time. In order to estimate the effect of popularity information on temporal exploration dynamics, we evaluate the average chain autocorrelation $\bar{\rho}(\tau)$ as follows:

\begin{equation}
    \bar{\rho}(\tau) = \dfrac{\sum_{c,t} \bm{x_{c,t}}' \cdot \bm{x_{c,t+\tau}'}}{\sum_{c,t} \bm{x_{c,t}}' \cdot
\bm{x_{c,t}}'  }
\end{equation}
where $\bm{x}_{c,t}'=\bm{x}_{c,t}-\bm{\mu}$ is the position (in semantic space, cf. Fig. \ref{fig:evolution}) of the image from chain $c$ created at generation $t$  after subtracting the mean coordinates $\bm{\mu}\equiv \frac{1}{C\times T} \sum_{c,t} \bm{x}_{c,t}$, and the dot product $(\cdot)$ measures the similarity between any image-pair. For a given time lag $\tau$, the autocorrelation $\bar{\rho}(\tau)$ measures the relative similarity between images at time $t$ and $t+\tau$ within a chain. High autocorrelation values imply that it takes many generations to visit regions farther away in semantic space, such that it eventually takes longer to explore the underlying cultural landscape. Fig. \ref{fig:autocorrelation}
shows that popularity information increases autocorrelation, which means initial conditions have a longer-lasting impact on the exploration process. 

Finally, we hypothesized that feedback loops owing to popularity information may have detrimental effects on the quality of cultural innovation. Quality was measured by collecting appeal, aesthetics, and creativity scores between 1 and 9, for every image, among a new set of 486 human participants (Fig. \ref{fig:ratings}). While differences in ``appeal'' were small, during most of the experiment (gen. 0--47), images in condition PI are less aesthetic ($p=0.002$) and creative ($p<0.001$). Towards the end of the experiment (gen. 48--59), however, we find this trend to be reduced and/or reversed. While further data is required to confidently ascertain the robustness of this inversion ($p=0.14$ for appeal and $p=0.14$ also for aesthetics), a potential explanation is proposed at the end of the paper.

In what follows, we show that these effects of social feedback are associated with changes in both selection processes, namely how participants choose ideas to build upon, and creation processes, namely how participants innovate

\subsubsection{Social feedback and selection}

\begin{figure*}[p]
    \centering
    \includegraphics[width=0.95\linewidth]{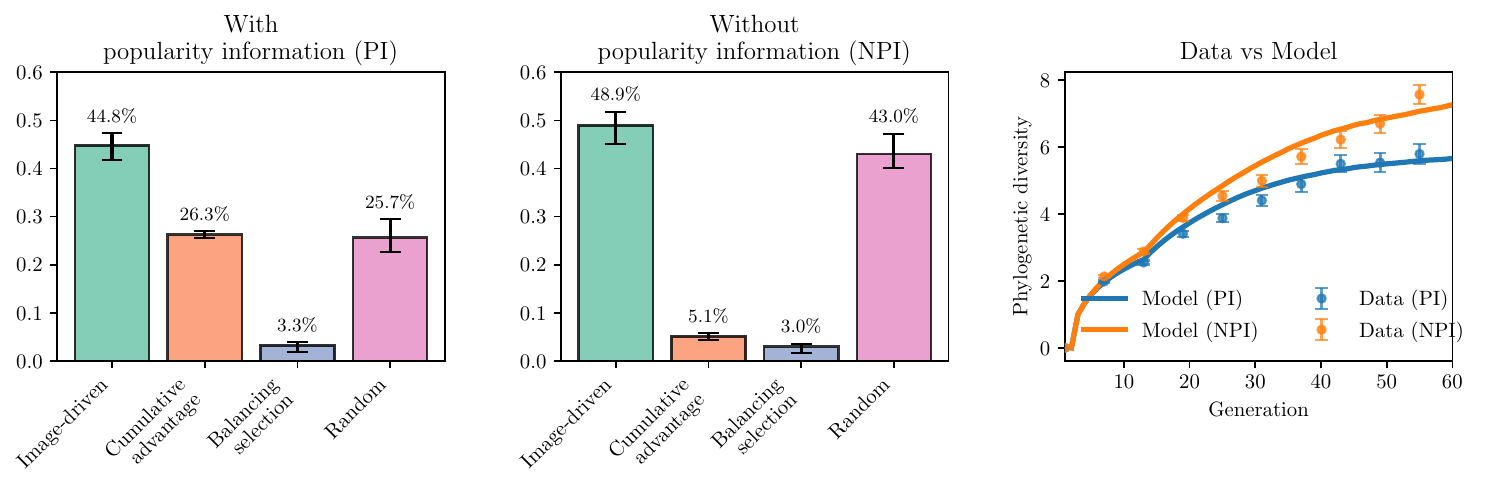}
    \caption{Left \& middle plots: effect of popularity information on selection. Right plot: phylogenetic diversity predicted by the best-fit mixture of selection strategies, for each condition. Error bars represent 68\% credible intervals (left \& middle plots) and standard errors  (right plot).}
    \label{fig:selection}
\end{figure*}

\begin{table*}[p]
\caption{Self-reported image selection strategies (clustered into 7 topics labeled with GPT-5.2), on a subset of the data (50\%). Columns indicate the number of participants in each cluster ($N$), the average prevalence of selection policies among these participants, and the distribution of these participants in each condition.}
\centering
\begin{tabular}{l|c|cccc|cc}
\toprule
Self-reported strategy cluster & $N$ & Image-driven & Cum. adv. & Bal. sel. & Random. & PI & NPI \\
\midrule
0\_Appealing and editable & $140$ & \cellcolor[HTML]{A4DBC9} 59\% & \cellcolor[HTML]{FEF0EA} 13\% & \cellcolor[HTML]{FCFDFD} 2\% & \cellcolor[HTML]{F8E0EF} 26\% & \cellcolor[HTML]{95BFDB} 47\% & \cellcolor[HTML]{FFBB7F} 53\% \\
1\_Editable improvement potential & $127$ & \cellcolor[HTML]{A3DAC9} 60\% & \cellcolor[HTML]{FEEFE9} 14\% & \cellcolor[HTML]{FFFFFF} 0\% & \cellcolor[HTML]{F8E0EF} 26\% & \cellcolor[HTML]{A7C9E1} 39\% & \cellcolor[HTML]{FFB06B} 61\% \\
2\_Popularity and past wins & $51$ & \cellcolor[HTML]{C6E8DD} 37\% & \cellcolor[HTML]{FDCDBB} 43\% & \cellcolor[HTML]{FDFEFE} 1\% & \cellcolor[HTML]{FAE8F3} 19\% & \cellcolor[HTML]{3E8ABE} 86\% & \cellcolor[HTML]{FFEDDD} 14\% \\
3\_No Deliberate Strategy & $42$ & \cellcolor[HTML]{A9DCCC} 56\% & \cellcolor[HTML]{FEF4F0} 9\% & \cellcolor[HTML]{FCFDFD} 2\% & \cellcolor[HTML]{F6D7EA} 34\% & \cellcolor[HTML]{9EC4DE} 43\% & \cellcolor[HTML]{FFB675} 57\% \\
4\_Popularity and editability & $48$ & \cellcolor[HTML]{A6DBCA} 58\% & \cellcolor[HTML]{FEE2D7} 25\% & \cellcolor[HTML]{FCFDFD} 2\% & \cellcolor[HTML]{FBECF5} 16\% & \cellcolor[HTML]{8FBBD9} 50\% & \cellcolor[HTML]{FFBF86} 50\% \\
5\_Aesthetic Appeal and Editability & $58$ & \cellcolor[HTML]{A6DBCA} 58\% & \cellcolor[HTML]{FEEAE2} 18\% & \cellcolor[HTML]{FDFEFE} 1\% & \cellcolor[HTML]{F9E4F1} 23\% & \cellcolor[HTML]{8FBBD9} 50\% & \cellcolor[HTML]{FFBF86} 50\% \\
6\_Gut appeal and potential & $20$ & \cellcolor[HTML]{A0D9C7} 62\% & \cellcolor[HTML]{FEF5F2} 8\% & \cellcolor[HTML]{F9FAFC} 5\% & \cellcolor[HTML]{F8E0EF} 26\% & \cellcolor[HTML]{83B4D5} 55\% & \cellcolor[HTML]{FFC592} 45\% \\
\bottomrule
\end{tabular}
\label{tab:topic_summary}
\end{table*}

We begin by exploring the impact of popularity information on selection. Using individual selection data, and for each condition (PI and NPI), we infer the prevalence of each of four selection strategies: image-driven selection (assuming choices are dictated by a latent intrinsic fitness); cumulative advantage (one of the most popular images is chosen); balancing selection (one of the least popular images is chosen); and uniformly random selection. In image-driven selection, we assume that participants select an image $k\sim\mathrm{Categorical}(\mathbf{p})$ where $\mathbf{p}=\mathrm{softmax}(\mathbf{u})$ and $\mathbf{u}$ is the latent utility of each image on the market. Latent utilities are assumed to balance four criteria $c$ (``appeal'', ``editing potential'', ``originality'', ``recognizability'') measured by latent scores $\mathbf{u}_c$. These scores are indirectly measured by collecting subjective ratings $r$  (from an additional new group of 413 participants) and assuming that $r_{ic}\sim\mathcal{N}(u_{ic},\sigma_c)$. We then further assume that $\mathbf{u}=\sum \beta_c \mathbf{u}_c$.

We fit this hierarchical model in Stan. This reveals a strong contribution of cumulative advantage in condition PI and its quasi-absence in condition NPI (Fig. \ref{fig:selection}). According to the model, balancing selection does not significantly arise; however, preferences for least popular variants could be mistaken as random selection. Overall, these observations are consistent with the reduction in diversity observed under social feedback. Interestingly, in condition PI, cumulative advantage arises at the expense of random selection, rather than image-driven selection (which does not significantly increase in condition NPI). Selection policy mixtures inferred from individual choices can be used to simulate the evolution of phylogenetic trees; this predicts lower phylogenetic under social feedback, consistent with the data. To further assess the robustness of our model, we analyzed the responses of participants in the original experiment, who were asked to self-report their selection strategy, and matched these responses to the mixture of policies inferred from their behavior. Using BERTopic \citep{grootendorst2022bertopic}, the self-reports were clustered into 7 topics, each of which was then  labeled with GPT-5.2. Table \ref{tab:topic_summary} shows that Topic 2, which emerges in condition PI and is dominated by reports referring to items' popularity, is strongly associated with cumulative advantage. 

Finally, like \citet{Salganik2006,Epstein2021}, we compared the inequality in the success of cultural items under the conditions with and without social feedback, in terms of the Gini coefficient \citep{bendel1989comparison}. We find $G(\textcolor{blue}{PI})=0.69$ and $G(\textcolor{orange}{NPI})=0.61$; in other words, in line with previous works, social feedback increases success inequalities ($\Delta=0.08, p < 0.001$).

\begin{figure}
    \centering
    \includegraphics[width=\linewidth]{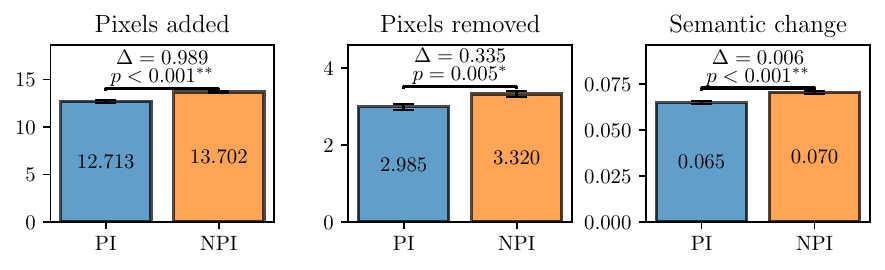}
    \caption{Average edits magnitude in each condition.}
    \label{fig:editing}
\end{figure}

\subsubsection{Social feedback and creation}

\begin{figure}
  \centering
    \includegraphics[width=\linewidth]{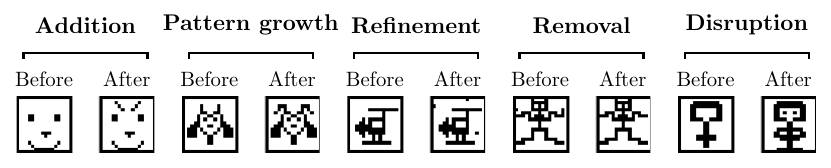}
    \caption{Illustrative examples for each of the five editing strategies.}
    \label{fig:examples}
\end{figure}

\begin{figure}
  \centering
  \includegraphics[width=0.97\linewidth]{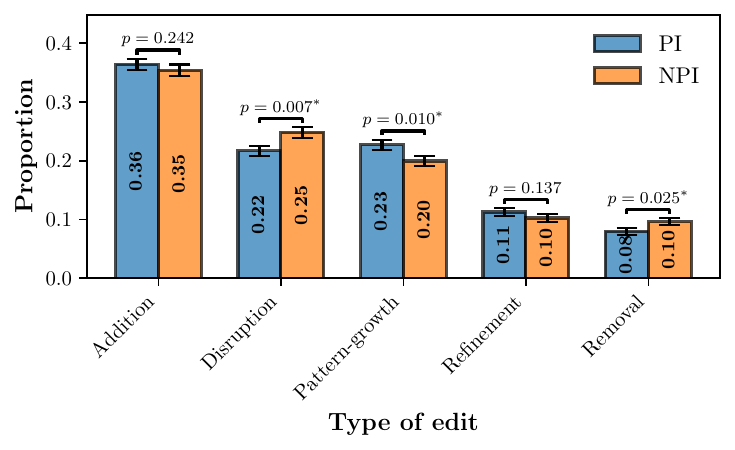}
  \caption{Prevalence of editing strategies across conditions. Error bars indicate 68\% CIs, under uniform priors. $p$ is the probability that the posterior includes the null (OR=1).}
  \label{fig:editing_strategies}
\end{figure}

During creation, under social feedback, participants change fewer pixels to the images they select and make smaller semantic shifts  on average (Fig. \ref{fig:editing}). However, summary statistics such as pixels added and removed say little about the underlying strategy. To get further insight into how popularity information affects the creation process itself, we classified edits into five categories (Fig. \ref{fig:examples}): ``disruption'' (when the new image represents something fundamentally different); ``addition'' (when visual features are added without changing the meaning of the image); ``pattern-growth'' (when existing features are expanded); ``removal'' (when existing features are removed, without changing the meaning of the image); and ``refinement'' (any other minor type of change). A new group of 372 humans were recruited online to classify parent-child image pairs into either of these categories, based on the majority-choice among three annotations per pair. We then calculated the prevalence of each strategy across conditions (Fig. \ref{fig:editing_strategies}). We find that in the condition with popularity information, participants are indeed less likely to disrupt the original idea (OR=$0.84, \text{CI}_{95\%}= [0.73, 0.96]$). Instead, participants are more likely to grow and expand existing visual features (OR=$1.18, \text{CI}_{95\%}= [1.02, 1.36]$), in line with the runaway hypothesis. Evidence also suggests that social feedback makes the removal of visual features less likely (OR=$0.82, \text{CI}_{95\%}= [0.66, 1.00]$).

\subsubsection{Aesthetics and popularity feedback}

\begin{figure}
    \centering
    \includegraphics[width=0.48\linewidth]{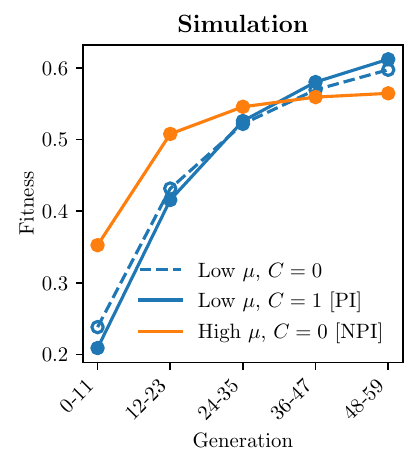}
    \includegraphics[width=0.5\linewidth]{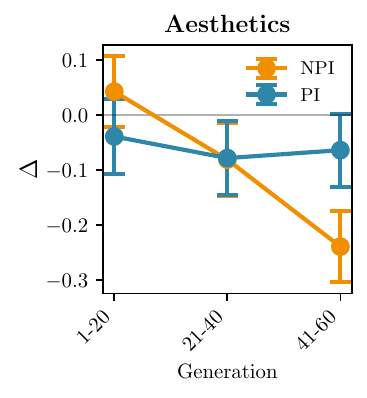}%
    \hfill
    
    \caption{Left: evolution of images' fitness in an idealized selection/mutation model, under low versus high mutation rates, and in the occurrence of cumulative advantage. Social feedback (PI) is captured by low mutation rates and cumulative advantage (low $\mu$, $C=1$), versus high mutation rates and no cumulative advantage for NPI (high $\mu$, $C=0$). Right: aesthetic improvements across generations and conditions in the data. $\Delta$ is the difference between the aesthetic scores of edited images and their parents. }
    \label{fig:improvements}
\end{figure}

Finally, we discuss why images might possibly be initially less but ultimately more aesthetic under social feedback (explaining the inversion seemingly occurring in the last iterations, cf. Fig. \ref{fig:ratings}). We consider a model where innovations are represented by bitstrings $(b_i)\in \{0,1\}^N$ with fitness $f=\sum_i b_i$ \citep[p.~269]{mackay2003information}. At each round, with 50\% probability, individuals select the image with highest fitness in the market (consistent with the measured rate of image-driven selection, Fig. \ref{fig:selection}), or otherwise choose at random. Then, they transform the image by mutating a few pixels. For higher mutation rates, as in condition NPI, the model predicts that images improve faster in the first generations (Fig. \ref{fig:improvements}). This is because when images are far from optimal, most mutations are beneficial. However, high mutation rates fare worse at later generations, when most changes are detrimental. The data indeed shows initially better but gradually worse variations in aesthetics in condition NPI, consistent with this model (Fig. \ref{fig:improvements}). Social feedback may thus have a protective effect that initially impedes improvements but ultimately preserves quality. While differences in mutation rates alone may explain an early disadvantage/late advantage under social feedback, cumulative advantage (present in condition PI) reinforces this effect (cf. the dashed versus blue solid lines in Fig. \ref{fig:improvements}).

\section{Discussion}

Our findings demonstrate that positive feedback loops in culture have distinct effects on selection and innovation, with measurable consequences for the diversity and quality of the cultural goods. Selection is distinctively influenced, with significant occurrence of cumulative advantage under social feedback, which ultimately reduces the diversity of cultural markets. By studying an open-ended creative task, we were additionally able to observe the diversity of innovation strategies and their alteration under the influence of social feedback: innovations were more conservative/meaning-preserving, and more likely to expand existing features, consistent with a cultural runaway process.

Future work, however, should address several limitations. First, control experiments should determine whether changes in innovation strategies partly follow from changes in the distribution of images being selected, which could be achieved by ablation studies similar to those from \citet{marjieh2025characterizing}. Second, participants did not observe the success of their own images; it is conceivable that they would adjust their strategies otherwise. Third, the constraint on editing, which intents to limit individual contributions, may not suffice to capture cumulative cultural evolution \citep{Miton2018}. Finally, in spatially extended systems, partial coupling between neighboring markets may alleviate or amplify some effects of social feedback, which could be addressed in a network-based experiment \citep{shiiku2025dynamics}. More broadly, our work demonstrates how innovations in large-scale online simulations of cultural evolution can inform our understanding of emergent features in cultural markets.


\section{Acknowledgments}

This work was supported by the NSF grant ``Collaborative Research: Designing smart environments to augment collective learning \& creativity'' (award 171418).

\nocite{MacCallum2012}
\printbibliography

\end{document}